\documentclass[aps,pra,twocolumn,superscriptaddress,groupedaddress,showpacs]{revtex4-1}
\usepackage{graphicx}
\usepackage{amsmath}
\usepackage{color}

\begin{document}

\title{Field Induced Oscillation of Two Majorana Modes for a finite Quantum Wire}
\author{Yue Yu}
\author{Kwok Yip Szeto}
\affiliation{Department of Physics, Hong Kong University of Science and Technology, Hong Kong}
\date{\today}
\begin{abstract}
The evolution of quantum walk on a finite wire under a small increment of vector potential $\alpha$ can exhibit intrinsic quantum oscillation of the two topologically protected bound states corresponding to the Majorana modes. By tuning an external electric field corresponding to the addition of an $\alpha$ impulse at the end of each intrinsic period, the intrinsic oscillation is enveloped by a beat modulation with a longer period. This beat oscillation is useful in the preparation of decoherence-free qubit in trapped ion chain and may be observed in several experiments.
\end{abstract}
\maketitle

\section{Introduction}
The recent advances in discrete time quantum walk \cite{A4,A5,B1,B2,B3} have created many interesting research endeavors in its application, such as in quantum computation \cite{A6,A7,A9,A10,A11}. 
However, finding a qubit that is free from decoherence is a major challenge. With our increased understanding of topologically protected systems, we find some recent solutions for this challenging problem \citep{Nielsen, Nayak}. 
In the literature of discrete quantum walk\cite{B4,B5,B6,B7,B8,B9,B10}, the Kitaev model \citep{Kitaev} for quantum wire is one of the simplest systems that supports these topological phases, which are identified as Majorana boundary modes \citep{Kitagawa, Hotat,B11,B12,B13,B14,B15,B16}. 
In this paper we use the exact solution for these modes in quantum walk in one dimension \citep{Hotat} to compute the response of the quantum walker to the stepwise increment of vector potential. We find interesting field induced oscillation, which can be useful for the preparation of the decoherence-free qubit. Our calculations involve numerical solutions of a set of coupled nonlinear equations, whose solution can be found analytically under certain limiting cases of long chain and small increment of vector potential. The limit of validity of our approximate calculation turns out to be easily satisfied for most cases and thus provides a useful guideline for the experimentalists. 
For quantum walk in one dimension, such as in a finite wire of length N, we can find the low energy boundary modes that is symmetry protected Majorana bound states by setting parameters in the coin matrix properly \cite{Hotat}. 
In this paper, we are going to study the interaction of the two bound states, and provide a description of these states under a coin matrix with a new phase $\alpha$ that corresponds to the addition of a vector potential. 
In one-dimensional discrete quantum walk, the evolution of the wave function is given by the unitary transformation $ U = S \otimes C$, which composes of the coin operator $C$ and a shifting operator 
$S = |L\rangle{\langle}L|\otimes\sum|n\rangle{\langle}n+1|+|R\rangle{\langle}R|\otimes\sum|n\rangle{\langle}n-1|$. 
Here, $|L,R\rangle$ are the basis of the coin space, while $|n\rangle$ are the basis of the position space. Here we focus on the coin operator in the following form 
\begin{equation}
C=\left( \begin{array}{cc}
\cos{\theta}e^{-i\alpha} & \sin{\theta}e^{-i\alpha} \\
-\sin{\theta}e^{i\alpha} & \cos{\theta}e^{i\alpha} \end{array} \right).\label{Eq:1}
\end{equation}
If the coin matrix is the same at every position in the system, the energy eigenstates can be written in the form a set of plane waves due to the translational symmetry. By substituting the eigenstate into the unitary operator, we can find the dispersion relationship of E and k,
\begin{equation}
\begin{aligned}
\psi_{k}(n,t) = e^{-iEt+ikn}\left[\begin{array}{ccc}
a_k\\
b_k \end{array} \right], \ 
\cos(E)=\cos{\theta}\cos(q),\label{Eq:3}
\end{aligned}
\end{equation}
where $q=k-\alpha$. For a given energy level, there are two corresponded eigen-momentum; the wave function is the linear combination of these two plane waves \citep{Hotat2}. The eigenvector $[a_k\,b_k]^{T}$ is
\begin{equation}
\left[\begin{array}{c}
a_k \\
b_k\end{array} \right]= 
\frac{1}{\sqrt{2-2\cos\theta\cos(E+q)}}\left[\begin{array}{c}
\sin\theta \\
(e^{-i(E+q)}-\cos\theta)\end{array} \right].\label{Eq:4}
\end{equation}
We are interested in the effect of a step-wise increase of the parameter $\alpha$ as a function of time, which corresponds to the step-wise increase of the vector potential applied to the one-dimensional system \citep{Hotat}. Experimentally this corresponds to the application of an impulse of electric field at specified time. Here k is the canonical momentum and $q$ is the kinetic momentum. 
For a given system with translational invariance and known eigenstate $\psi_{\alpha}$ for a given $\alpha$, there exists a simple relation between the eigenstate if $\alpha$ is changed to $\alpha'$ uniformly for every points in the system as, $\psi_{\alpha'}(n,t)=\psi_{\alpha}(n,t)e^{i(\alpha'-\alpha)n }$. 
The relative phase between the two neighbouring points will compensate the effect of $\alpha'-\alpha$ in the coin matrix, so that
\begin{equation}
\begin{aligned}
&(C_{\alpha'}\psi_{\alpha'}(n,t))_L=\psi_{\alpha'}(n-1,t+1)_L,
\\&(C_{\alpha'}\psi_{\alpha'}(n,t))_R=\psi_{\alpha'}(n+1,t+1)_R.
\end{aligned}\label{Eq:6}
\end{equation}
This implies that the eigen-energy and the parameter $q$ are unchanged, but the quasi-momentum $k=q+\alpha$ is replaced by $k'=q+\alpha'$. This result is important when we tune the $\alpha$ in a step-wise manner. 
 
\section{Evolution of the bound states under single impulse}
We first summarize the known results for the bound state that appears at the topological boundary where the rotation parameter $\theta$ changes its signs. For the system with single boundary, the ground state eigen-energy is exactly zero. Now we consider a finite system of length N+2; the coordinates are set to be $n=0...N+1$. Let us consider the example with the following setup of the finite chain: $\theta_1$ is negative at both ends $n=0,N+1$ while $\theta_0$ is a positive constant in the middle. In this way, we have two topological boundaries, and we anticipate the existence of two symmetry protected bound states\citep{Kitagawa, Kitaev, Hotat}. Specifically, if we set $\theta_1=-\pi/2$ so that the diagonal terms in the coin matrix in Eq.\ref{Eq:1} is zero, then the motion of the walker will be completely reflected at the boundaries. Physically this describes quantum walk on a finite wire.
Now we consider the bound state of this system. Due to the change of the signs of $\theta$, the wave function is delocalized at the two ends for this ground state. Equivalently, two quasi-particles are formed around $n=1$ and $n=N$. The eigen-energy $E_0$ of the ground state can deviate from zero, as there can be an effective interaction between them\cite{Hotat}. 
We first review our exact solution of these two bound states for a zero $\alpha$. 
Physically, if there were boundary modes, their interaction energy should depend on the size of the system as well as the coin parameter $\theta$, which describes the inertia of the walker. We propose an ansatz for the wave function of the ground state as a linear combination of two momentum eigenstates,
\begin{equation}
\begin{aligned}
\psi_{k}(n,t)= &e^{-iEt}(c_{L}e^{-\kappa{n}}\left[\begin{array}{ccc}a_{i\kappa}\\b_{i\kappa} \end{array} \right]+\\&c_{R}e^{-\kappa{(N+1-n)}}\left[\begin{array}{ccc}a_{-i\kappa}\\b_{-i\kappa} \end{array} \right]) ,\,n=1...N.
\end{aligned}\label{Eq:7}
\end{equation}
For the bound state, the momentum is purely imaginary, so we set $\kappa=-ik$ for simplicity. The eigenvector $[a_{\pm{i}\kappa},b_{{\pm}i\kappa}]^{T}$ is given by Eq.\ref{Eq:4}. At the two ends, the wave function needs to satisfy the continuity requirement,
\begin{equation}
\begin{split}
\psi_{k}(0,t) = e^{-iEt}[c_{L}a_{i\kappa}+c_{R}e^{-\kappa{(N+1)}}a_{-i\kappa}, 0]^T, \\
\psi_{k}(N+1,t) = e^{-iEt}[0, c_{L}e^{-\kappa{(N+1)}}b_{i\kappa}+c_{R}b_{-i\kappa}]^T.
\end{split}
\label{Eq:8}
\end{equation}
The boundary conditions at the two ends provide two linear equations about $c_L$ and $c_R$.
By solving the equation for the determinant of the coefficients matrix, we can get solutions for $(\kappa,E)$. 
We first solve this numerically and we find that the bound state energy is very small, thus we assume the small E limit and obtain analytically the following dispersion relation:
\begin{equation}
e^{\kappa}=\frac{1+\sin\theta}{\cos\theta}+O(E^2).\label{Eq:11}
\end{equation}
In this small E approximation, we get the expression of the energy $E_0$,
\begin{equation}
E_0={\pm}2\tan\theta_0(\frac{\cos\theta_0}{1+\sin\theta_0})^{N+1}.\label{Eq:13}
\end{equation}
With the approximate solutions for $\kappa$ and $E_0$ , we can proceed to compute analytically the change of the system under the application of an impulse realized by a stepwise increase of the parameter $\alpha$. 

The corresponding boundary mode with energy $+E_0$ evolves as
\begin{equation}
\begin{split}
\psi_{0}^{+}(n,t) = &e^{-iE_{0}t} \bigg( c_{L}e^{-\kappa{n}}\left[\begin{array}{ccc}a_{i\kappa}\\b_{i\kappa} \end{array} \right]+\\&c_{R}e^{-\kappa{(N+1-n)}}\left[\begin{array}{ccc}a_{-i\kappa}\\b_{-i\kappa} \end{array} \right] \bigg) ,\,n=1...N.
\end{split}\label{Eq:14}
\end{equation}
Here the superscript denotes energy of the mode, in this case it is the mode with energy $+E_0$ . (The other mode has energy $-E_0$.) The subscript denotes the vector potential parameter $\alpha$ and in this case it is 0. 

At time $t=0^+$, we add a non-zero ${\alpha_0}$ impulse to the entire wire, then the original wave function can be expressed as a linear combination of the set of new eigenstates of the system after the impulse with the nonzero ${\alpha_0}$, $\psi(n,0)_{0}^{+}=c_{+}\psi^{+}_{{\alpha_0}}(n,0)+c_{-}\psi^{-}_{{\alpha_0}}(n,0)+R$, 
where $ \psi^{\pm}_{\alpha_0}$ are the two ground states after adding ${\alpha_0}$ and the term $R=\sum_{i\neq{\pm}}{c_i}\psi^{i}_{\alpha_0}$ describes the projection to all those states that are not the boundary modes. If we assume that $R$ is small, then we can describe the evolution of the system after the addition of the nonzero vector potential $\alpha_0$ by a new state with a two-level system composed only with the two ground states $ \psi^{\pm}_{\alpha_0}$. 
We will later verify that this two-level approximation is valid for a small step-wise change of $\alpha$. Using Eq.\ref{Eq:6}, we can write the new eigenstates as, $\psi^{\pm}_{{\alpha_0}}(n,t)=\psi^{\pm}_{0}(n,t)e^{i{\alpha_0}{n}}$ and then compute the coefficients $c_{\pm}$ by the inner products between the old and new eigenstates. 
\begin{equation}
\begin{split}
c_{\pm}&=\sum_{n}\psi_0^{+}(n,0)\psi_{{\alpha_0}}^{\pm}(n,0)^{*} \\&=\sum_{n}\psi_{0}^{+}(n,0)^{*}\psi_{0}^{\pm}(n,0)e^{-i{\alpha_0}{n}}.
\label{Eq:17}
\end{split}
\end{equation}
Note that the new wave function will oscillate since it is no longer a stationary state after the impulse. 
The frequencies of the oscillation modes are equal to the energy gaps between different energy eigenstates of the system with $\alpha_o >0$. 
In our two-level system approximation, the only energy gap is $2E_0$. The error of this approximation can be estimated by looking at the effect of neglecting R, which is the inner product of $\psi_0^{+}(n,0)$ with other higher energy levels. 
This error can be estimated by calculating the total probability of $|c_{+}|^2+|c_{-}|^2$, which should be close to 1 if the higher energy levels terms in R can be ignored. 

\begin{figure}[h]
\centering
\includegraphics[width=4.2cm]{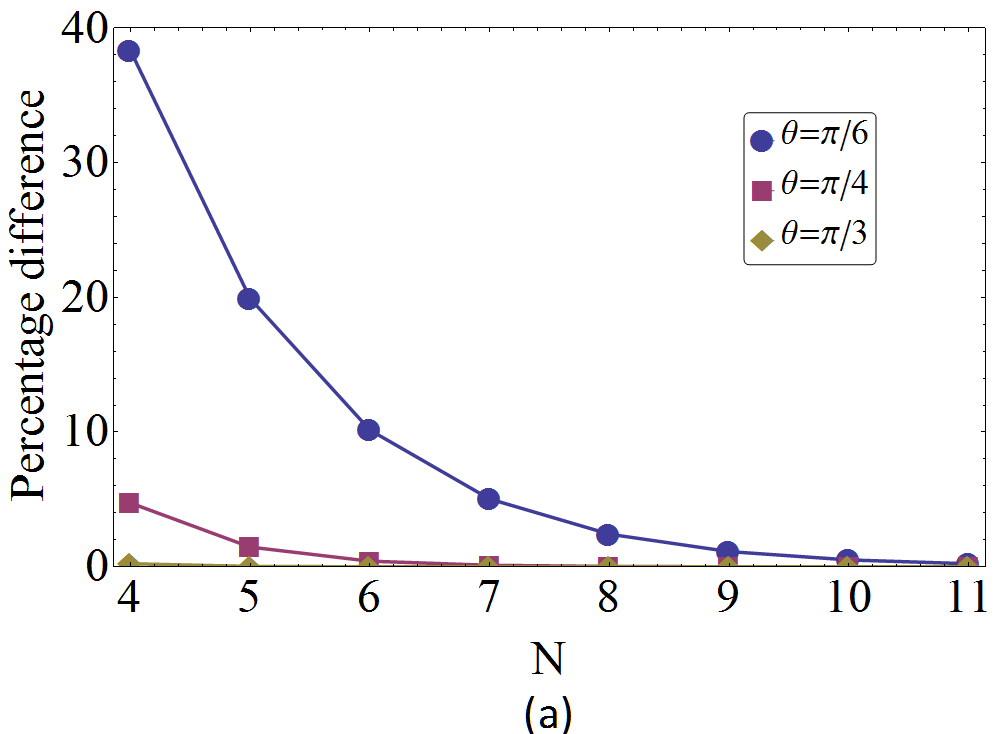}
\includegraphics[width=4.2cm]{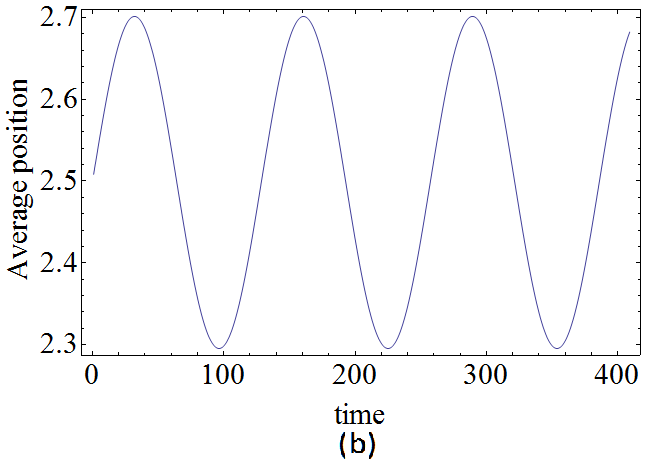}
\caption{(a) The percentage difference of the slope in Eq.\ref{Eq:20}. The error is determined by N and $\theta$. For $\theta=\pi/4$, which corresponds to the Hadamard walk, the error is smaller than 5$\%$ for N=4. (b) The intrinsic oscillation between the two ground states. The size of the system is N=4. An impulse ${\alpha_0}=0.01\pi$ is added at $t=0$. The intrinsic oscillation $T_0$ is around 128.5.}\label{Fig 1}
\end{figure}

We first consider the case of long chain for which our approximate expression for $\kappa$ and $E_0$ are very good \cite{Hotat}. Physically, for a long wire, the two quasi-particles are effectively isolated. Under this condition, $\psi_0^{+}(n,0)^{*}\psi_{{\alpha_0}}^{\pm}(n,0)$ can be rewritten as four exponential functions,
\begin{equation}
\begin{aligned}
&\psi_{0}^{+}(n,0)^{*}\psi_{0}^{\pm}(n,0)_L =A(e^{-2\kappa{n}}{\pm}e^{-2\kappa(N-n)}), \\
&\psi_{0}^{+}(n+1,0)^{*}\psi_{0}^{\pm}(n+1,0)_R=A(e^{-2\kappa{n}}{\pm}e^{-2\kappa(N-n)})\\
&n=0...N.
\end{aligned}\label{Eq:18}
\end{equation}
The first two terms are from the left component and the last two terms are from the right component of the wave function. 
The $\pm$ sign come from the orthogonality of the original two ground states. 
For a long wire, $N\gg{1}$, it is reasonable to let the upper limit in the summand in Eq.\ref{Eq:17} to go to infinity, $n\rightarrow\infty$, thereby obtaining the analytical result,
\begin{equation}
c_{\pm}=\frac{1-e^{-2\kappa}}{4}(\frac{1}{1-e^{-2\kappa-i{\alpha_0}}}\pm\frac{e^{-iN{\alpha_0}}}{1-e^{-2\kappa+i{\alpha_0}}})(1+e^{-i{\alpha_0}}).\label{Eq:19}
\end{equation}
If now we apply a very small impulse so that $N{\alpha_0}\ll{1}$, then we can expand $c_{+}$ and $c_{-}$ by keeping terms up to $O({\alpha_0}^2)$,
\begin{widetext}
\begin{equation}
\begin{aligned}
& c_{+}=1-(\frac{N+1}{2})i{\alpha_0}-(\dfrac{1}{4}N^2+\frac{N}{4}-\frac{N}
{2(e^{2\kappa}-1)}+\frac{1}{4}+\frac{e^{2\kappa}+1}{2(e^{2\kappa}-1)^2}){\alpha_0}^2, \\
& c_{-}=-(\frac{1}{e^{2\kappa}-1}-\frac{N}{2})i{\alpha_0}+(\dfrac{1}{4}N^2+\frac{N}{4}-\frac{N+1}{2(e^{2\kappa}-1)}){\alpha_0}^2, \\
& |c_{+}|^2+|c_{-}|^2=1-(\frac{1}{4}+\frac{e^{2\kappa}}{(e^{2\kappa}-1)^2}){\alpha_0}^2.
\end{aligned}\label{Eq:20}
\end{equation}
\end{widetext}
When $N{\alpha_0}\ll{1}$, the total probability becomes linear in ${\alpha_0}^2$ only. 
However, even when N is small, the above expression for $ c_{\pm}$ is still good as long as $N{\alpha_0}\ll{1}$. 
This can be seen from a comparison of this simple expression in Eq.\ref{Eq:20} with the numerical calculation without ignoring R for different $\theta$ at N=4 ( Fig.\ref{Fig 1}).
We see that the percentage difference for the Hadamard walk for N=4 is only 5$\%$ for a single impulse that change $\alpha$ from 0 to $\alpha_0=0.01\pi$. 
The condition for the two-level system approximation to be good is thus $N{\alpha_0}\ll{1}$. 
To understand the physics of our two-level approximation, let’s consider a small N (=4) system. 
The ground state energy is $7.78\times{10}^{-3}\pi$, while the first excited state has energy $0.34\pi$. 
Thus, we expect that the impulse on the system by switching from zero to a small $\alpha_0\ll{1\over N}$ does not excite the system to higher energy levels through the effective E field produced by the sudden change in vector potential, since the energy gap $\delta E =2E_o$ between the two ground states $E_0$ and $-E_0$ is much smaller than any other energy gaps. We can therefore observe the intrinsic oscillation of the two level system (two Majorana modes) in a finite wire as long as our change in vector potential is very small ($\alpha_0\ll{1/N}$) .

\section{Evolution of bound states under multiple impulses}
The characteristic period of the intrinsic oscillation of the two-level system with energy difference $\delta E=2E_0$ is $T_0=2\pi/\delta E=\pi/E_o$. This can be observed by taking some time average of an observable over $T_0$. A more interesting oscillation is related to the increase of the vector potential ${\alpha}$ by hitting the system with multiple impulses so that 
$\alpha(t)=m\alpha_0,\,t\in[(m-1)T_0,mT_0),\,m=0,1,2...$ with a small $\alpha_0$ such that $N\alpha_0\ll{1}$. 
From the analysis of single impulse, we expect that it is still reasonable to neglect other states R in the projection if $N\alpha_0\ll{1}$, 
so that the wave function can be written as a linear combination of the two ground states and the two-level approximation is still good. 
\begin{equation}
\begin{aligned}
\psi(n,t)&=c_{+}e^{-iE_0t} \psi^{+}_{\alpha}(n,t)+c_{-} e^{iE_0t} \psi^{-}_{\alpha}(n,t).\label{Eq:23}
\end{aligned}
\end{equation}
At the end of each intrinsic period $t=mT_0$, a new $\alpha$ is introduced; and the wave function is projected to a new basis of eigenfunctions. During the time interval $[(m-1)T_0,mT_0)$, both eigenstates $ \psi^{\pm}_{\alpha}$ go through half of period $\pi/E_0$, and the wave function before the next projection is the same state as $t=(m-1)T_0$ after the previous projection,
\begin{equation}
\begin{aligned}
&\psi(n,mT^{-}_0)=-(c_{+} \psi^{+}_{\alpha}(n)+c_{-} \psi^{-}_{\alpha}(n)),\\
&\psi(n,(m-1)T^{+}_0)=c_{+}\psi^{+}_{\alpha}(n)+c_{-} \psi^{-}_{\alpha}(n).
\end{aligned}\label{Eq:24}
\end{equation}
As a result, the two ground states do not change between subsequent projections, except for the overall minus sign. 
As long as $\alpha_0$ is small, the projection planes are close to each other. 
Although those states in R, which are outside of our two-level system, are changed in the time interval between impulses, they have higher energy and their projection are oscillating with a much higher frequency, so that the contribution of R is still very small as subsequent projections are close, which is true when $\alpha_0$ is small. 
Thus, we can use our two-level approximation to compute the time evolution of the two coefficient $c_{+}$ and $c_{-}$ analytically as in the case of single impulse.  
Because we are interested in the oscillations larger than the intrinsic period, we can measure time by the number of intrinsic periods $m$. 
For each impulse, the transition matrix $P_{\alpha}$ between states in the two level system can be defined as, 
$[c_{+},c_{-}]^T_{m+1}=P_{\alpha}[c_{+},c_{-}]^T_{m}$.
The elements in the $2\times{2}$ transition matrix $P_{\alpha}$ are the inner products between the new and old basis in the two-level system,
\begin{equation}
\begin{split}
P_{ij}&=\sum_{n}\psi^{i}_{\alpha+\alpha_0}(n)^{*}\psi^{j}_{\alpha}(n)
=\sum_{n}\psi^{i}_{\alpha_0}(n)^{*}\psi^{j}_{0}(n),\, i,j=\pm.\label{Eq:27}
\end{split}
\end{equation}
Note that $P_{ij}(\alpha)$ depends on $\alpha_0$, since the phase $e^{i\alpha{n}}$ of the new and old basis cancels. We thus have
\begin{equation}
\begin{aligned}
P_{++}& =P_{--}=1-(\frac{N+1}{2})i\alpha_0\\&-(\dfrac{1}{4}N^2+\frac{N}{4}-\frac{N}{2(e^{2\kappa}-1)}+\frac{1}{4}+\frac{e^{2\kappa}+1}{2(e^{2\kappa}-1)^2})\alpha_0^2,\\
 P_{+-}&=P_{-+}=-(\frac{1}{e^{2\kappa}-1}-\frac{N}{2})i\alpha_0
\\&+(\dfrac{1}{4}N^2+\frac{N}{4}-\frac{N+1}{2(e^{2\kappa}-1)})\alpha_0^2.
\end{aligned}\label{Eq:28}
\end{equation}
and the eigenvalues and eigenvectors of the transition matrix are,
$\lambda_{1,2}=P_{++}\pm{P}_{+-},\,
v_{1,2}=[\frac{1}{\sqrt{2}},\pm\frac{1}{\sqrt{2}}]^{T}$.

The absolute values of these two eigenvalues are less than one, implying a loss of information. Further calculations show that they have equal magnitude, $|\lambda_{1,2}|^2=1-(\frac{1}{4}+\frac{e^{2\kappa}}{(e^{2\kappa}-1)^2})\alpha_0^2$. We see that the total probability decreases with a speed proportional to $\alpha_0^2$ after each intrinsic period, so that after $m$ intrinsic periods, the total probability becomes 
\begin{equation}
\begin{aligned}
|c_{+}|^2+|c_{-}|^2&=(|\lambda_{1,2}|^2)^m\\&=
\left( 1-\left( \frac{1}{4}+\frac{e^{2\kappa}}{(e^{2\kappa}-1)^2}\right) \alpha_0^2\right) ^m.
\end{aligned}\label{Eq:31}
\end{equation}
This expression gives the evolution of the total probability using the two-level system approximation when the system is under the application of impulse $m$ times, each time the value of $\alpha$ is increased by $\alpha_0$ and the impulse is applied periodically with an interval $T_0$. 
For the same value of $N\alpha_0$, the longer the wire, the smaller is the loss of probability, and the better is the two level system approximation.
\begin{figure}[htb]
\includegraphics[width=6.5cm]{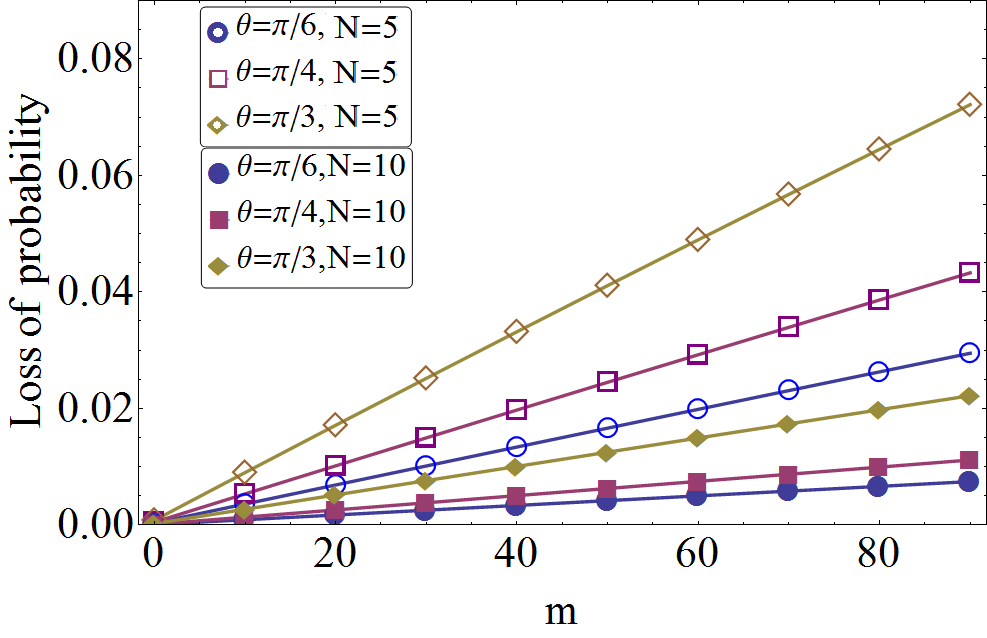}
\caption{The loss of probability under multiple impulses. Three different $\theta$ are used on the N=5,10 system. $\alpha_0$ is set to be $0.01\pi$ and $0.005\pi$ respectively so that in both cases, $N\alpha_0 =0.05\pi$. The loss of probability is less than 5$\%$ for Hadamard coin after m=100 intrinsic oscillations. 
The prediction is Eq. \ref{Eq:31}, shown here by the lines.}\label{Fig 2}
\end{figure}

\begin{figure}[htb]
\centering
\includegraphics[width=4.2cm]{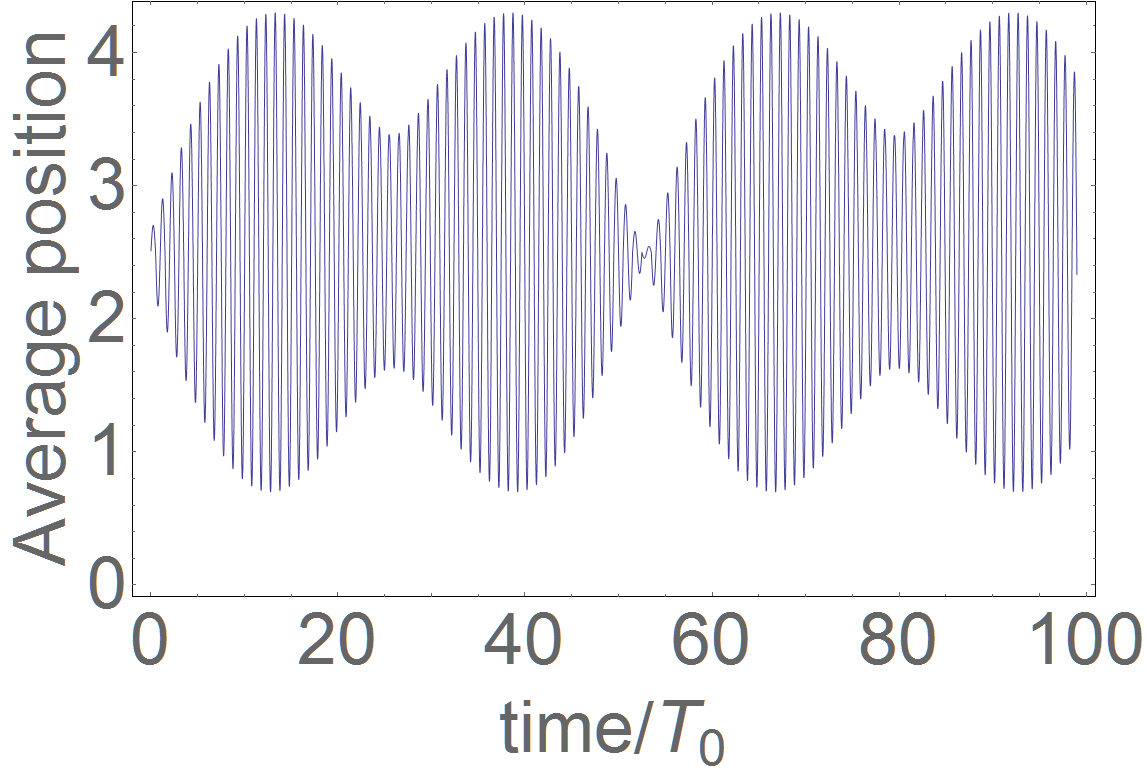}
\includegraphics[width=4.35cm]{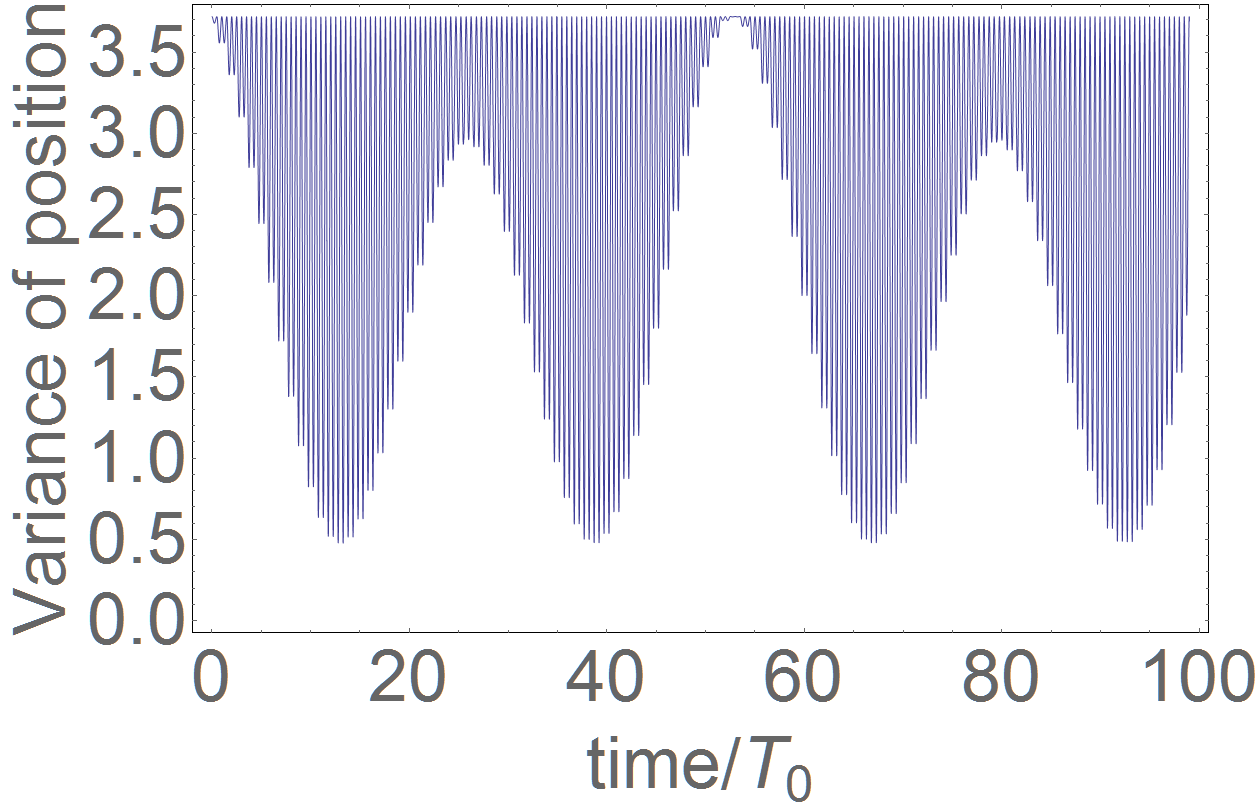}
\caption{The average position and variance of Hadamard walk with periodic impulse. Only the two ground states are considered and the total probability is normalized. The size of the wire is N=4, and the unit impulse is $\alpha_0=0.01\pi$. The system evolves for m=100 intrinsic periods. Note the system returns to original state $E=E_0$
around $m=\frac{2\pi}{\tan^{-1}(\Delta\phi)}=55.8$.}\label{Fig 3}
\end{figure}

Our numerical results (Fig.\ref{Fig 2}) show that for Hadamard coin with $N\alpha_0 =0.05\pi$, the loss of probability is less than 5$\%$ for $N=5$ after 100 intrinsic oscillations. 
In Fig.\ref{Fig 3} we show the average position and variance of the Hadamard walk under the application of multiple impulse. We see the intrinsic oscillation is enclosed by an envelope function with a longer period. The phase difference $\Delta\phi$ is exactly the beat frequency of this two level system, $f_{\text{beat}}=\Delta\phi=(N-\frac{2}{e^{2\kappa}-1})\alpha_0$ with the period of the beat being $T_{\text{beat}}/T_0=\frac{2\pi}{\Delta\phi}$. 
\section{Observation of Quantum Oscillation}
\begin{figure}[htb]
\centering
\includegraphics[width=4cm]{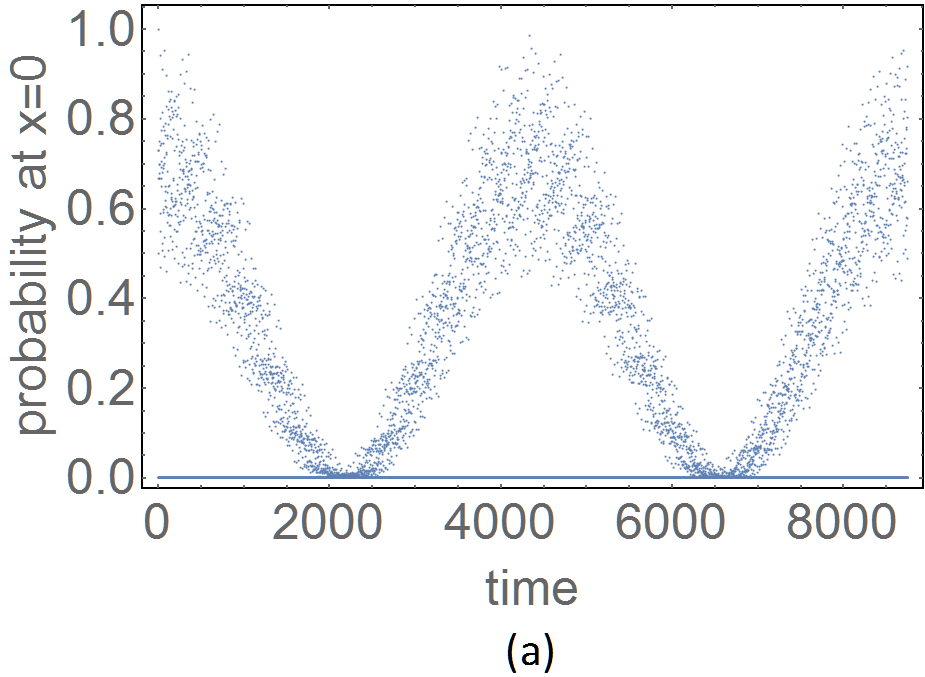}
\includegraphics[width=4.1cm]{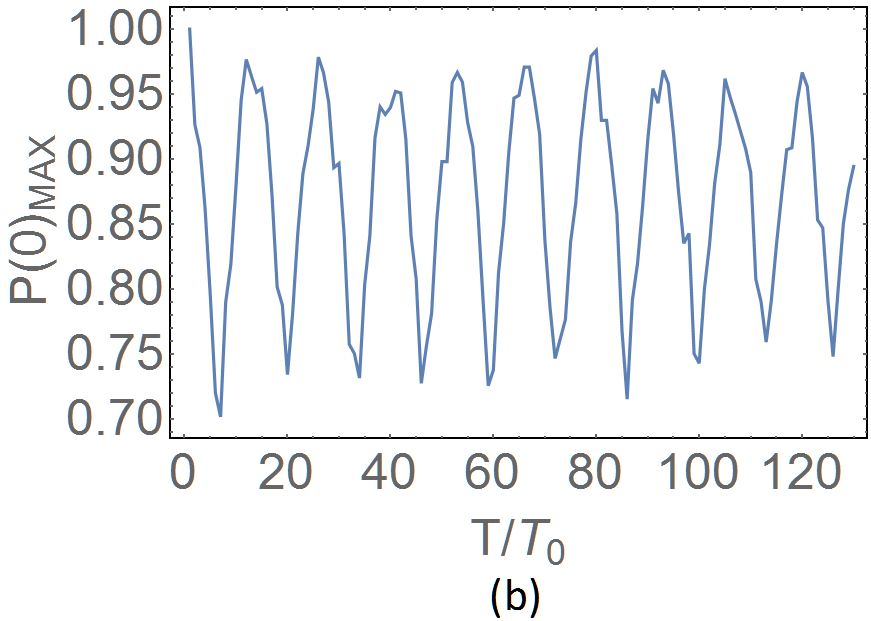}
\caption{(a) The probability at $n=0$ in the system with length $N+2=10$, for Hadamard walk. Here the intrinsic oscillation is clearly observed in the first two intrinsic periods, which is $T_0\approx4376$. (b) The beat oscillation in the quantum wire can be seen from the maximum probability at $n=0$ at each intrinsic period $T_{0}$. Here the length of the wire is $N+2=10$ for Hadamard walk and the vector potential parameter is $\alpha_0=0.02\pi$. }\label{Fig 4}
\end{figure}

\begin{figure}[htb]
\centering
\includegraphics[width=8cm]{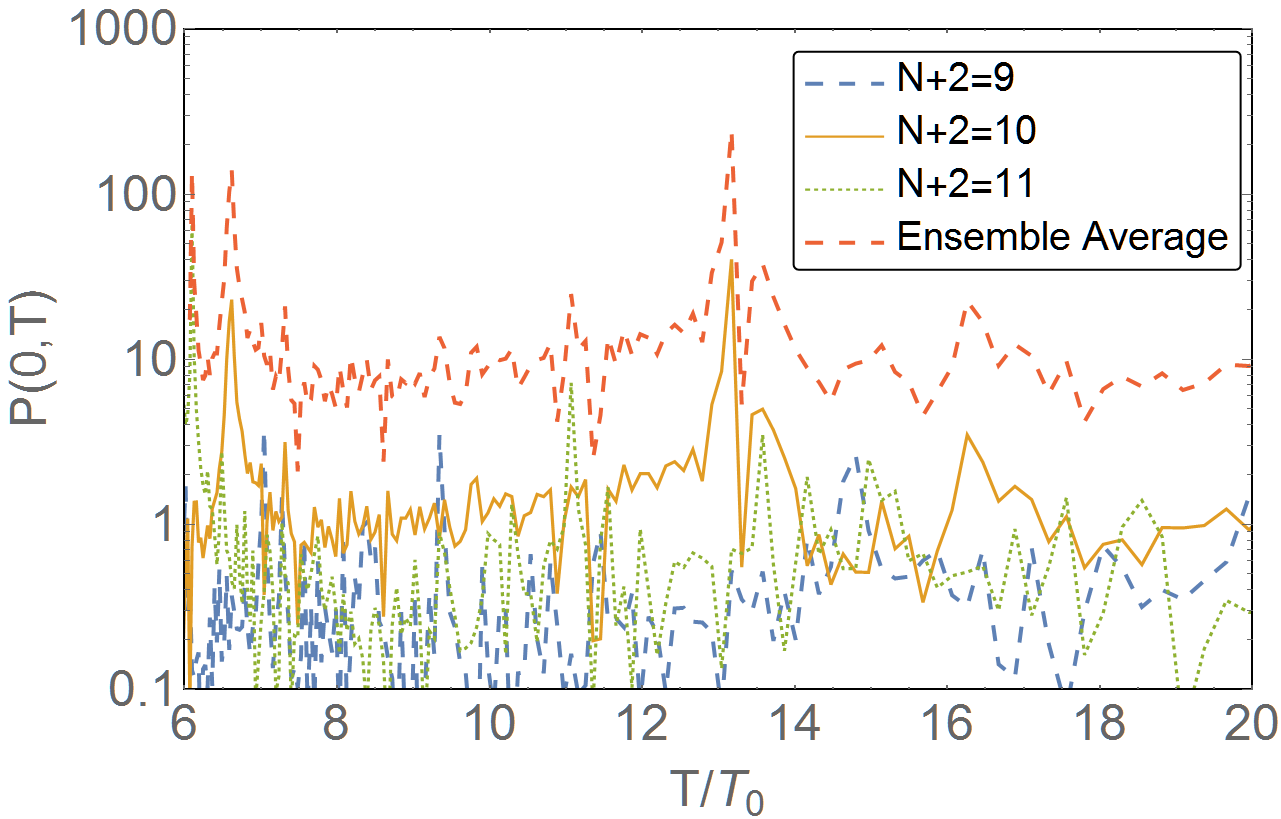}
\caption{The evolution of the probability at $n=0$ when we add a small increment of $\alpha_0=0.02\pi$ to the ensemble of wires of length $N+2= 9, 10, 11$ with corresponding concentration of $0.2, 0.6, 0.2$. We observe that only the correct system can provide a peak of the beat oscillation at $13.2T_0$. The ensemble average is rescaled by ten.
}\label{Fig 8}
\end{figure}

The intrinsic oscillation originated from the energy gap between the two Majorana modes can be observed if we start the quantum walker at the end of the wire ($n=0$). 
This initial state $\psi(0,0)_{L}=1$ can be approximated by a linear combination of the two Majorana modes in the form of Eq.\ref{Eq:14}, since all the other scattering states are not localized. In Fig. \ref{Fig 4}(a), we show the numerical results of the quantum walk starting with this initial state and the result can be compared with a measurement of the probability distribution at $n=0$ as a function of time. Furthermore, when we apply a small increase of the external electric field at the end of each intrinsic period, the numerical results show the beat oscillation in Fig. \ref{Fig 4}(b), which can be compared with the measurement of the maximum probability of the quantum walker at $n=0$ after each intrinsic period $T_o$. 
Now, for a real system of quantum wires, their lengths usually cannot be all equal. 
We therefore generalize our numerical calculation to an ensemble of quantum wires of different lengths arranged in parallel along the $x$ direction, with the left end points of these wires forming a surface in the $yz$ plane. If the initial state of the walkers is a linear combination of the Majorana modes with maximum probability at $x=0$ on this surface, we can observe the beat oscillation on the wire with a particular length $N_0$ by adding an electric field at the end of the intrinsic period $T_0=2\pi/\delta E=\pi/E_o$ where $E_o(N_0)$ is computed for that specific length $N_0$. 
To illustrate this selection mechanism on the beat oscillation, let us consider an ensemble of wires of length $N+2= 9, 10, 11$ with corresponding concentration of $0.2, 0.6, 0.2$. Numerical calculation for this ensemble is shown in Fig. \ref{Fig 8}. 
The three sets of beat pattern are superimposed and the beat period for the ensemble is $12.5T_0$ from simulation, 
which is very close to the value $13.2 T_0$ for a pure sample with all wires of length $N+2=10$ 

\section{Discussion}
In this paper, we first set the two topological boundary conditions at the ends of a finite wire to construct the two quasi-particles which are shown to be Majorana modes recently \citep{Hotat, Kitaev,Kitagawa}. From the exact calculation of these modes\citep{Hotat}, we obtain the interaction between the two quasi-particles and show the deviation of the ground state energy $\pm {E_o}$, computed for a quantum wire of specific length $N_0$.  In order to explore the quantum oscillation of these modes, we initialize the system in the ground state ($c_{+}=1, c_{-}=0$) and apply a single electric impulse at $T_{0}$ by an increase of the vector potential parameter $\alpha_0$ in the coin matrix. 
Before the application of the impulse, the variance of the intrinsic oscillation is at a maximum because the wave function is delocalized, while the average position is in the middle of the wire. 
After the addition of an impulse, the wave function starts to oscillate with a frequency equal to the energy gap $2E_0$ in our two-level approximation which is valid when $\alpha_0$ is sufficiently less than $1 /N$ .  
The oscillating probability density $|\psi|^2$, its associated probability current, and the average speed are all physical observables. 
This field induced oscillation provides a possible way to realize the delocalized ground state, with a characteristic intrinsic period $T_{0}=\pi / E_o$. 
We also consider the application of multiple impulse through the increase of the potential parameter $\alpha_0$ in the coin matrix, at designated time $T_m=mT_{0}$. 
Numerical calculation shows that there is an envelope function describable as a beat oscillation. 
For example, if the walker starts with zero velocity at the end of the wire of length 4, the wave function will evolve to become the wave package at $t=14T_0$. 
By slowly increasing the coin parameter $\alpha$ at $t=mT_0$ by the small amount $\alpha_0$ (as used in Fig. \ref{Fig 3}), then the delocalized ground state is expected to appear around $t=56T_0$, with very limited mixture with the other states. The variance of probability distribution function goes to minimum (see Fig.\ref{Fig 3}), which means that the wave function is similar to a narrow wave package at $t=\frac{T_{\text{beat}}}{4}\approx 14T_0$, while its oscillation amplitude is maximum. 
In principle, this wave function can go to another delocalized ground state ($c_{+}=0, |c{-}|=1$) around $t=\frac{T_{\text{beat}}}{2}\approx28T_0$. 
However, this does not happen due to interference because for $N=4$, the intrinsic period is $T_0\approx128.5$, which is not an integer. 
For $N=3$, where $T_0$ is closer to an integer, the system can reach the other delocalized state clearly. This observation may be useful in the preparation of robust qubit by choosing appropriate N and $\theta_0$ so that the intrinsic period $T_0$ is very close to an integer. 
Note that one may reduce the beat period by increasing $\alpha_0$, since the beat frequency is proportional to $\alpha_0$ in lowest order. 
However, there is a trade-off between the reduction of the beat period and the goodness of the approximation of the two-level system that requires a small $\alpha_0$.  
In conclusion, this field induced oscillation provides a possible way to prepare the delocalized ground state.

We expect the one-dimensional quantum walk on finite wire can be realized in a trapped-ion chain with an Majorana qubit encoded \citep{Mezzacapo}. 
This qubit is topologically protected against major sources of decoherence, thereby providing an efficient quantum memory. 
The oscillation of these symmetry protected bound states provide a new phenomenon that can be exploited. 
Experimentally, we can place the finite wire in $xy$ plane, add electric field along $x$ in a stepwise manner with period $T_0$, 
then our theory predicts the observation of the oscillation on the average position of the particle with a specified beat frequency. 
This long period beat oscillation is particularly impressive compared with relatively fast intrinsic oscillation due to the nature of discrete quantum walk. 
For the $Ca^{2+}$ realization of quantum walk, the hopping time is in the order of $2ns$, while the beat period for $N=4$ system is with the order of $20\mu{s}$, with significant improved coherence lifetimes. 
A second experiment where our results can be tested is through the electrical measurements on indium antimonide nanowires contacted with two electrodes, one normal and one superconducting \cite{black1}. In these experiments, the electron density can be changed by varying the gate voltage. By properly measuring the probability current, one may be able to observe our predicted beat oscillation when the electric field is increased at a rate defined by the intrinsic frequency $T_0$ which in turn is fixed by the length of the nanowires. 
A third experiment that may test our prediction is the work by \cite{black2}. 
By placing ferromagnetic atomic chains on the surface of superconducting lead, they fabricated a one-dimensional topological superconductor with the appearance Majorana fermions similar to our model of one-dimensional quantum walk on a finite wire. 
Other recent works on black phosphorus \cite{wangning} may be another example for the detection of the beat oscillation. These experimental systems may directly provide evidence of the Majorana modes. 

\begin{acknowledgments}
We acknowledge discussion with Lam Hotat and Wang Ning.
\end{acknowledgments}
\bibliography{bound}

\begin{thebibliography}{33}%
\makeatletter
\providecommand \@ifxundefined [1]{%
 \@ifx{#1\undefined}
}%
\providecommand \@ifnum [1]{%
 \ifnum #1\expandafter \@firstoftwo
 \else \expandafter \@secondoftwo
 \fi
}%
\providecommand \@ifx [1]{%
 \ifx #1\expandafter \@firstoftwo
 \else \expandafter \@secondoftwo
 \fi
}%
\providecommand \natexlab [1]{#1}%
\providecommand \enquote  [1]{``#1''}%
\providecommand \bibnamefont  [1]{#1}%
\providecommand \bibfnamefont [1]{#1}%
\providecommand \citenamefont [1]{#1}%
\providecommand \href@noop [0]{\@secondoftwo}%
\providecommand \href [0]{\begingroup \@sanitize@url \@href}%
\providecommand \@href[1]{\@@startlink{#1}\@@href}%
\providecommand \@@href[1]{\endgroup#1\@@endlink}%
\providecommand \@sanitize@url [0]{\catcode `\\12\catcode `\$12\catcode
  `\&12\catcode `\#12\catcode `\^12\catcode `\_12\catcode `\%12\relax}%
\providecommand \@@startlink[1]{}%
\providecommand \@@endlink[0]{}%
\providecommand \url  [0]{\begingroup\@sanitize@url \@url }%
\providecommand \@url [1]{\endgroup\@href {#1}{\urlprefix }}%
\providecommand \urlprefix  [0]{URL }%
\providecommand \Eprint [0]{\href }%
\providecommand \doibase [0]{http://dx.doi.org/}%
\providecommand \selectlanguage [0]{\@gobble}%
\providecommand \bibinfo  [0]{\@secondoftwo}%
\providecommand \bibfield  [0]{\@secondoftwo}%
\providecommand \translation [1]{[#1]}%
\providecommand \BibitemOpen [0]{}%
\providecommand \bibitemStop [0]{}%
\providecommand \bibitemNoStop [0]{.\EOS\space}%
\providecommand \EOS [0]{\spacefactor3000\relax}%
\providecommand \BibitemShut  [1]{\csname bibitem#1\endcsname}%
\let\auto@bib@innerbib\@empty
\bibitem [{\citenamefont {Meyer}(1996)}]{A4}%
  \BibitemOpen
  \bibfield  {author} {\bibinfo {author} {\bibfnamefont {D.}~\bibnamefont
  {Meyer}},\ }\href {\doibase 10.1007/BF02199356} {\bibfield  {journal}
  {\bibinfo  {journal} {Journal of Statistical Physics}\ }\textbf {\bibinfo
  {volume} {85}},\ \bibinfo {pages} {551} (\bibinfo {year} {1996})}\BibitemShut
  {NoStop}%
\bibitem [{\citenamefont {Kempe}(2003)}]{A5}%
  \BibitemOpen
  \bibfield  {author} {\bibinfo {author} {\bibfnamefont {J.}~\bibnamefont
  {Kempe}},\ }\href {\doibase 10.1080/00107151031000110776} {\bibfield
  {journal} {\bibinfo  {journal} {Contemporary Physics}\ }\textbf {\bibinfo
  {volume} {44}},\ \bibinfo {pages} {307} (\bibinfo {year} {2003})}\BibitemShut
  {NoStop}%
\bibitem [{\citenamefont {Feynman}(1948)}]{B1}%
  \BibitemOpen
  \bibfield  {author} {\bibinfo {author} {\bibfnamefont {R.~P.}\ \bibnamefont
  {Feynman}},\ }\href {\doibase 10.1103/RevModPhys.20.367} {\bibfield
  {journal} {\bibinfo  {journal} {Rev. Mod. Phys.}\ }\textbf {\bibinfo {volume}
  {20}},\ \bibinfo {pages} {367} (\bibinfo {year} {1948})}\BibitemShut
  {NoStop}%
\bibitem [{\citenamefont {Feynman}(1986)}]{B2}%
  \BibitemOpen
  \bibfield  {author} {\bibinfo {author} {\bibfnamefont {R.~P.}\ \bibnamefont
  {Feynman}},\ }\href {\doibase 10.1007/BF01886518} {\bibfield  {journal}
  {\bibinfo  {journal} {Foundations of Physics}\ }\textbf {\bibinfo {volume}
  {16}},\ \bibinfo {pages} {507} (\bibinfo {year} {1986})}\BibitemShut
  {NoStop}%
\bibitem [{\citenamefont {Aharonov}\ \emph {et~al.}(1993)\citenamefont
  {Aharonov}, \citenamefont {Davidovich},\ and\ \citenamefont {Zagury}}]{B3}%
  \BibitemOpen
  \bibfield  {author} {\bibinfo {author} {\bibfnamefont {Y.}~\bibnamefont
  {Aharonov}}, \bibinfo {author} {\bibfnamefont {L.}~\bibnamefont
  {Davidovich}}, \ and\ \bibinfo {author} {\bibfnamefont {N.}~\bibnamefont
  {Zagury}},\ }\href {\doibase 10.1103/PhysRevA.48.1687} {\bibfield  {journal}
  {\bibinfo  {journal} {Phys. Rev. A}\ }\textbf {\bibinfo {volume} {48}},\
  \bibinfo {pages} {1687} (\bibinfo {year} {1993})}\BibitemShut {NoStop}%
\bibitem [{\citenamefont {Lovett}\ \emph {et~al.}(2010)\citenamefont {Lovett},
  \citenamefont {Cooper}, \citenamefont {Everitt}, \citenamefont {Trevers},\
  and\ \citenamefont {Kendon}}]{A6}%
  \BibitemOpen
  \bibfield  {author} {\bibinfo {author} {\bibfnamefont {N.~B.}\ \bibnamefont
  {Lovett}}, \bibinfo {author} {\bibfnamefont {S.}~\bibnamefont {Cooper}},
  \bibinfo {author} {\bibfnamefont {M.}~\bibnamefont {Everitt}}, \bibinfo
  {author} {\bibfnamefont {M.}~\bibnamefont {Trevers}}, \ and\ \bibinfo
  {author} {\bibfnamefont {V.}~\bibnamefont {Kendon}},\ }\href {\doibase
  10.1103/PhysRevA.81.042330} {\bibfield  {journal} {\bibinfo  {journal} {Phys.
  Rev. A}\ }\textbf {\bibinfo {volume} {81}},\ \bibinfo {pages} {042330}
  (\bibinfo {year} {2010})}\BibitemShut {NoStop}%
\bibitem [{\citenamefont {Childs}\ \emph {et~al.}(2013)\citenamefont {Childs},
  \citenamefont {Gosset},\ and\ \citenamefont {Webb}}]{A7}%
  \BibitemOpen
  \bibfield  {author} {\bibinfo {author} {\bibfnamefont {A.~M.}\ \bibnamefont
  {Childs}}, \bibinfo {author} {\bibfnamefont {D.}~\bibnamefont {Gosset}}, \
  and\ \bibinfo {author} {\bibfnamefont {Z.}~\bibnamefont {Webb}},\ }\href
  {\doibase 10.1126/science.1229957} {\bibfield  {journal} {\bibinfo  {journal}
  {Science}\ }\textbf {\bibinfo {volume} {339}},\ \bibinfo {pages} {791}
  (\bibinfo {year} {2013})}\BibitemShut {NoStop}%
\bibitem [{\citenamefont {Childs}\ \emph {et~al.}(2003)\citenamefont {Childs},
  \citenamefont {Cleve}, \citenamefont {Deotto}, \citenamefont {Farhi},
  \citenamefont {Gutmann},\ and\ \citenamefont {Spielman}}]{A9}%
  \BibitemOpen
  \bibfield  {author} {\bibinfo {author} {\bibfnamefont {A.~M.}\ \bibnamefont
  {Childs}}, \bibinfo {author} {\bibfnamefont {R.}~\bibnamefont {Cleve}},
  \bibinfo {author} {\bibfnamefont {E.}~\bibnamefont {Deotto}}, \bibinfo
  {author} {\bibfnamefont {E.}~\bibnamefont {Farhi}}, \bibinfo {author}
  {\bibfnamefont {S.}~\bibnamefont {Gutmann}}, \ and\ \bibinfo {author}
  {\bibfnamefont {D.~A.}\ \bibnamefont {Spielman}},\ }in\ \href {\doibase
  10.1145/780542.780552} {\emph {\bibinfo {booktitle} {Proceedings of the
  Thirty-fifth Annual ACM Symposium on Theory of Computing}}},\ \bibinfo
  {series and number} {STOC '03}\ (\bibinfo  {publisher} {ACM},\ \bibinfo
  {address} {New York, NY, USA},\ \bibinfo {year} {2003})\ pp.\ \bibinfo
  {pages} {59--68}\BibitemShut {NoStop}%
\bibitem [{\citenamefont {Shenvi}\ \emph {et~al.}(2003)\citenamefont {Shenvi},
  \citenamefont {Kempe},\ and\ \citenamefont {Whaley}}]{A10}%
  \BibitemOpen
  \bibfield  {author} {\bibinfo {author} {\bibfnamefont {N.}~\bibnamefont
  {Shenvi}}, \bibinfo {author} {\bibfnamefont {J.}~\bibnamefont {Kempe}}, \
  and\ \bibinfo {author} {\bibfnamefont {K.~B.}\ \bibnamefont {Whaley}},\
  }\href {\doibase 10.1103/PhysRevA.67.052307} {\bibfield  {journal} {\bibinfo
  {journal} {Phys. Rev. A}\ }\textbf {\bibinfo {volume} {67}},\ \bibinfo
  {pages} {052307} (\bibinfo {year} {2003})}\BibitemShut {NoStop}%
\bibitem [{\citenamefont {Ambainis}\ \emph {et~al.}(2005)\citenamefont
  {Ambainis}, \citenamefont {Kempe},\ and\ \citenamefont {Rivosh}}]{A11}%
  \BibitemOpen
  \bibfield  {author} {\bibinfo {author} {\bibfnamefont {A.}~\bibnamefont
  {Ambainis}}, \bibinfo {author} {\bibfnamefont {J.}~\bibnamefont {Kempe}}, \
  and\ \bibinfo {author} {\bibfnamefont {A.}~\bibnamefont {Rivosh}},\ }in\
  \href {http://dl.acm.org/citation.cfm?id=1070432.1070590} {\emph {\bibinfo
  {booktitle} {Proceedings of the Sixteenth Annual ACM-SIAM Symposium on
  Discrete Algorithms}}},\ \bibinfo {series and number} {SODA '05}\ (\bibinfo
  {publisher} {Society for Industrial and Applied Mathematics},\ \bibinfo
  {address} {Philadelphia, PA, USA},\ \bibinfo {year} {2005})\ pp.\ \bibinfo
  {pages} {1099--1108}\BibitemShut {NoStop}%
\bibitem [{\citenamefont {Nielsen}\ and\ \citenamefont
  {Chuang}(2000)}]{Nielsen}%
  \BibitemOpen
  \bibfield  {author} {\bibinfo {author} {\bibfnamefont {M.~A.}\ \bibnamefont
  {Nielsen}}\ and\ \bibinfo {author} {\bibfnamefont {I.~L.}\ \bibnamefont
  {Chuang}},\ }\href@noop {} {\emph {\bibinfo {title} {Quantum Computation and
  Quantum Information}}},\ \bibinfo {edition} {1st}\ ed.\ (\bibinfo
  {publisher} {Cambridge University Press},\ \bibinfo {address} {Cambridge},\
  \bibinfo {year} {2000})\BibitemShut {NoStop}%
\bibitem [{\citenamefont {Nayak}\ \emph {et~al.}(2008)\citenamefont {Nayak},
  \citenamefont {Simon}, \citenamefont {Stern}, \citenamefont {Freedman},\ and\
  \citenamefont {Das~Sarma}}]{Nayak}%
  \BibitemOpen
  \bibfield  {author} {\bibinfo {author} {\bibfnamefont {C.}~\bibnamefont
  {Nayak}}, \bibinfo {author} {\bibfnamefont {S.~H.}\ \bibnamefont {Simon}},
  \bibinfo {author} {\bibfnamefont {A.}~\bibnamefont {Stern}}, \bibinfo
  {author} {\bibfnamefont {M.}~\bibnamefont {Freedman}}, \ and\ \bibinfo
  {author} {\bibfnamefont {S.}~\bibnamefont {Das~Sarma}},\ }\href {\doibase
  10.1103/RevModPhys.80.1083} {\bibfield  {journal} {\bibinfo  {journal} {Rev.
  Mod. Phys.}\ }\textbf {\bibinfo {volume} {80}},\ \bibinfo {pages} {1083}
  (\bibinfo {year} {2008})}\BibitemShut {NoStop}%
\bibitem [{\citenamefont {Ambainis}\ \emph {et~al.}(2001)\citenamefont
  {Ambainis}, \citenamefont {Bach}, \citenamefont {Nayak}, \citenamefont
  {Vishwanath},\ and\ \citenamefont {Watrous}}]{B4}%
  \BibitemOpen
  \bibfield  {author} {\bibinfo {author} {\bibfnamefont {A.}~\bibnamefont
  {Ambainis}}, \bibinfo {author} {\bibfnamefont {E.}~\bibnamefont {Bach}},
  \bibinfo {author} {\bibfnamefont {A.}~\bibnamefont {Nayak}}, \bibinfo
  {author} {\bibfnamefont {A.}~\bibnamefont {Vishwanath}}, \ and\ \bibinfo
  {author} {\bibfnamefont {J.}~\bibnamefont {Watrous}},\ }in\ \href {\doibase
  10.1145/380752.380757} {\emph {\bibinfo {booktitle} {Proceedings of the
  Thirty-third Annual ACM Symposium on Theory of Computing}}},\ \bibinfo
  {series and number} {STOC '01}\ (\bibinfo  {publisher} {ACM},\ \bibinfo
  {address} {New York, NY, USA},\ \bibinfo {year} {2001})\ pp.\ \bibinfo
  {pages} {37--49}\BibitemShut {NoStop}%
\bibitem [{\citenamefont {{Nayak}}\ and\ \citenamefont
  {{Vishwanath}}(2000)}]{B5}%
  \BibitemOpen
  \bibfield  {author} {\bibinfo {author} {\bibfnamefont {A.}~\bibnamefont
  {{Nayak}}}\ and\ \bibinfo {author} {\bibfnamefont {A.}~\bibnamefont
  {{Vishwanath}}},\ }\href@noop {} {\bibfield  {journal} {\bibinfo  {journal}
  {eprint arXiv:quant-ph/0010117}\ } (\bibinfo {year} {2000})},\ \Eprint
  {http://arxiv.org/abs/quant-ph/0010117} {quant-ph/0010117} \BibitemShut
  {NoStop}%
\bibitem [{\citenamefont {Bach}\ \emph {et~al.}(2004)\citenamefont {Bach},
  \citenamefont {Coppersmith}, \citenamefont {Goldschen}, \citenamefont
  {Joynt},\ and\ \citenamefont {Watrous}}]{B6}%
  \BibitemOpen
  \bibfield  {author} {\bibinfo {author} {\bibfnamefont {E.}~\bibnamefont
  {Bach}}, \bibinfo {author} {\bibfnamefont {S.}~\bibnamefont {Coppersmith}},
  \bibinfo {author} {\bibfnamefont {M.~P.}\ \bibnamefont {Goldschen}}, \bibinfo
  {author} {\bibfnamefont {R.}~\bibnamefont {Joynt}}, \ and\ \bibinfo {author}
  {\bibfnamefont {J.}~\bibnamefont {Watrous}},\ }\href {\doibase
  http://dx.doi.org/10.1016/j.jcss.2004.03.005} {\bibfield  {journal} {\bibinfo
   {journal} {Journal of Computer and System Sciences}\ }\textbf {\bibinfo
  {volume} {69}},\ \bibinfo {pages} {562 } (\bibinfo {year}
  {2004})}\BibitemShut {NoStop}%
\bibitem [{\citenamefont {Kendon}(2007)}]{B7}%
  \BibitemOpen
  \bibfield  {author} {\bibinfo {author} {\bibfnamefont {V.}~\bibnamefont
  {Kendon}},\ }\href {\doibase 10.1017/S0960129507006354} {\bibfield  {journal}
  {\bibinfo  {journal} {Mathematical. Structures in Comp. Sci.}\ }\textbf
  {\bibinfo {volume} {17}},\ \bibinfo {pages} {1169} (\bibinfo {year}
  {2007})}\BibitemShut {NoStop}%
\bibitem [{\citenamefont {Chandrashekar}\ \emph {et~al.}(2008)\citenamefont
  {Chandrashekar}, \citenamefont {Srikanth},\ and\ \citenamefont
  {Laflamme}}]{B8}%
  \BibitemOpen
  \bibfield  {author} {\bibinfo {author} {\bibfnamefont {C.~M.}\ \bibnamefont
  {Chandrashekar}}, \bibinfo {author} {\bibfnamefont {R.}~\bibnamefont
  {Srikanth}}, \ and\ \bibinfo {author} {\bibfnamefont {R.}~\bibnamefont
  {Laflamme}},\ }\href {\doibase 10.1103/PhysRevA.77.032326} {\bibfield
  {journal} {\bibinfo  {journal} {Phys. Rev. A}\ }\textbf {\bibinfo {volume}
  {77}},\ \bibinfo {pages} {032326} (\bibinfo {year} {2008})}\BibitemShut
  {NoStop}%
\bibitem [{\citenamefont {Romanelli}(2009)}]{B9}%
  \BibitemOpen
  \bibfield  {author} {\bibinfo {author} {\bibfnamefont {A.}~\bibnamefont
  {Romanelli}},\ }\href {\doibase 10.1103/PhysRevA.80.042332} {\bibfield
  {journal} {\bibinfo  {journal} {Phys. Rev. A}\ }\textbf {\bibinfo {volume}
  {80}},\ \bibinfo {pages} {042332} (\bibinfo {year} {2009})}\BibitemShut
  {NoStop}%
\bibitem [{\citenamefont {Shikano}\ and\ \citenamefont {Katsura}(2010)}]{B10}%
  \BibitemOpen
  \bibfield  {author} {\bibinfo {author} {\bibfnamefont {Y.}~\bibnamefont
  {Shikano}}\ and\ \bibinfo {author} {\bibfnamefont {H.}~\bibnamefont
  {Katsura}},\ }\href {\doibase 10.1103/PhysRevE.82.031122} {\bibfield
  {journal} {\bibinfo  {journal} {Phys. Rev. E}\ }\textbf {\bibinfo {volume}
  {82}},\ \bibinfo {pages} {031122} (\bibinfo {year} {2010})}\BibitemShut
  {NoStop}%
\bibitem [{\citenamefont {{Kitaev}}(2001)}]{Kitaev}%
  \BibitemOpen
  \bibfield  {author} {\bibinfo {author} {\bibfnamefont {A.~Y.}\ \bibnamefont
  {{Kitaev}}},\ }\href {\doibase 10.1070/1063-7869/44/10S/S29} {\bibfield
  {journal} {\bibinfo  {journal} {Physics Uspekhi}\ }\textbf {\bibinfo {volume}
  {44}},\ \bibinfo {pages} {131} (\bibinfo {year} {2001})}\BibitemShut
  {NoStop}%
\bibitem [{\citenamefont {Kitagawa}\ \emph {et~al.}(2012)\citenamefont
  {Kitagawa}, \citenamefont {Broome}, \citenamefont {Fedrizzi}, \citenamefont
  {Rudner}, \citenamefont {Berg}, \citenamefont {Kassal}, \citenamefont
  {Aspuru-Guzik}, \citenamefont {Demler},\ and\ \citenamefont
  {White}}]{Kitagawa}%
  \BibitemOpen
  \bibfield  {author} {\bibinfo {author} {\bibfnamefont {T.}~\bibnamefont
  {Kitagawa}}, \bibinfo {author} {\bibfnamefont {M.~A.}\ \bibnamefont
  {Broome}}, \bibinfo {author} {\bibfnamefont {A.}~\bibnamefont {Fedrizzi}},
  \bibinfo {author} {\bibfnamefont {M.~S.}\ \bibnamefont {Rudner}}, \bibinfo
  {author} {\bibfnamefont {E.}~\bibnamefont {Berg}}, \bibinfo {author}
  {\bibfnamefont {I.}~\bibnamefont {Kassal}}, \bibinfo {author} {\bibfnamefont
  {A.}~\bibnamefont {Aspuru-Guzik}}, \bibinfo {author} {\bibfnamefont
  {E.}~\bibnamefont {Demler}}, \ and\ \bibinfo {author} {\bibfnamefont {A.~G.}\
  \bibnamefont {White}},\ }\href {\doibase 10.1038/ncomms1872} {\bibfield
  {journal} {\bibinfo  {journal} {Nature Communications}\ }\textbf {\bibinfo
  {volume} {3}},\ \bibinfo {pages} {882} (\bibinfo {year} {2012})}\BibitemShut
  {NoStop}%
\bibitem [{\citenamefont {Lam}\ \emph {et~al.}(2015)\citenamefont {Lam},
  \citenamefont {Yu},\ and\ \citenamefont {Szeto}}]{Hotat}%
  \BibitemOpen
  \bibfield  {author} {\bibinfo {author} {\bibfnamefont {H.}~\bibnamefont
  {Lam}}, \bibinfo {author} {\bibfnamefont {Y.}~\bibnamefont {Yu}}, \ and\
  \bibinfo {author} {\bibfnamefont {K.~Y.}\ \bibnamefont {Szeto}},\ }\href
  {\doibase 10.1103/PhysRevA.93.052319} {\  (\bibinfo {year} {2015}),\
  10.1103/PhysRevA.93.052319}\BibitemShut {NoStop}%
\bibitem [{\citenamefont {Kitagawa}\ \emph {et~al.}(2010)\citenamefont
  {Kitagawa}, \citenamefont {Rudner}, \citenamefont {Berg},\ and\ \citenamefont
  {Demler}}]{B11}%
  \BibitemOpen
  \bibfield  {author} {\bibinfo {author} {\bibfnamefont {T.}~\bibnamefont
  {Kitagawa}}, \bibinfo {author} {\bibfnamefont {M.~S.}\ \bibnamefont
  {Rudner}}, \bibinfo {author} {\bibfnamefont {E.}~\bibnamefont {Berg}}, \ and\
  \bibinfo {author} {\bibfnamefont {E.}~\bibnamefont {Demler}},\ }\href
  {\doibase 10.1103/PhysRevA.82.033429} {\bibfield  {journal} {\bibinfo
  {journal} {Phys. Rev. A}\ }\textbf {\bibinfo {volume} {82}},\ \bibinfo
  {pages} {033429} (\bibinfo {year} {2010})}\BibitemShut {NoStop}%
\bibitem [{\citenamefont {Asb\'oth}\ and\ \citenamefont {Obuse}(2013)}]{B12}%
  \BibitemOpen
  \bibfield  {author} {\bibinfo {author} {\bibfnamefont {J.~K.}\ \bibnamefont
  {Asb\'oth}}\ and\ \bibinfo {author} {\bibfnamefont {H.}~\bibnamefont
  {Obuse}},\ }\href {\doibase 10.1103/PhysRevB.88.121406} {\bibfield  {journal}
  {\bibinfo  {journal} {Phys. Rev. B}\ }\textbf {\bibinfo {volume} {88}},\
  \bibinfo {pages} {121406} (\bibinfo {year} {2013})}\BibitemShut {NoStop}%
\bibitem [{\citenamefont {Asb\'oth}(2012)}]{B13}%
  \BibitemOpen
  \bibfield  {author} {\bibinfo {author} {\bibfnamefont {J.~K.}\ \bibnamefont
  {Asb\'oth}},\ }\href {\doibase 10.1103/PhysRevB.86.195414} {\bibfield
  {journal} {\bibinfo  {journal} {Phys. Rev. B}\ }\textbf {\bibinfo {volume}
  {86}},\ \bibinfo {pages} {195414} (\bibinfo {year} {2012})}\BibitemShut
  {NoStop}%
\bibitem [{\citenamefont {Cedzich}\ \emph {et~al.}(2016)\citenamefont
  {Cedzich}, \citenamefont {Grünbaum}, \citenamefont {Stahl}, \citenamefont
  {Velázquez}, \citenamefont {Werner},\ and\ \citenamefont {Werner}}]{B14}%
  \BibitemOpen
  \bibfield  {author} {\bibinfo {author} {\bibfnamefont {C.}~\bibnamefont
  {Cedzich}}, \bibinfo {author} {\bibfnamefont {F.~A.}\ \bibnamefont
  {Grünbaum}}, \bibinfo {author} {\bibfnamefont {C.}~\bibnamefont {Stahl}},
  \bibinfo {author} {\bibfnamefont {L.}~\bibnamefont {Velázquez}}, \bibinfo
  {author} {\bibfnamefont {A.~H.}\ \bibnamefont {Werner}}, \ and\ \bibinfo
  {author} {\bibfnamefont {R.~F.}\ \bibnamefont {Werner}},\ }\href
  {http://stacks.iop.org/1751-8121/49/i=21/a=21LT01} {\bibfield  {journal}
  {\bibinfo  {journal} {Journal of Physics A: Mathematical and Theoretical}\
  }\textbf {\bibinfo {volume} {49}},\ \bibinfo {pages} {21LT01} (\bibinfo
  {year} {2016})}\BibitemShut {NoStop}%
\bibitem [{\citenamefont {{Meyer}}(2001)}]{B15}%
  \BibitemOpen
  \bibfield  {author} {\bibinfo {author} {\bibfnamefont {D.~A.}\ \bibnamefont
  {{Meyer}}},\ }\href {\doibase 10.1088/0305-4470/34/35/323} {\bibfield
  {journal} {\bibinfo  {journal} {Journal of Physics A Mathematical General}\
  }\textbf {\bibinfo {volume} {34}},\ \bibinfo {pages} {6981} (\bibinfo {year}
  {2001})}\BibitemShut {NoStop}%
\bibitem [{\citenamefont {{Meyer}}(1998)}]{B16}%
  \BibitemOpen
  \bibfield  {author} {\bibinfo {author} {\bibfnamefont {D.~A.}\ \bibnamefont
  {{Meyer}}},\ }\href {\doibase 10.1088/0305-4470/31/10/009} {\bibfield
  {journal} {\bibinfo  {journal} {Journal of Physics A Mathematical General}\
  }\textbf {\bibinfo {volume} {31}},\ \bibinfo {pages} {2321} (\bibinfo {year}
  {1998})}\BibitemShut {NoStop}%
\bibitem [{\citenamefont {Lam}\ and\ \citenamefont {Szeto}(2015)}]{Hotat2}%
  \BibitemOpen
  \bibfield  {author} {\bibinfo {author} {\bibfnamefont {H.}~\bibnamefont
  {Lam}}\ and\ \bibinfo {author} {\bibfnamefont {K.~Y.}\ \bibnamefont
  {Szeto}},\ }\href {\doibase 10.1103/PhysRevA.92.012323} {\bibfield  {journal}
  {\bibinfo  {journal} {PhysRevA}\ }\textbf {\bibinfo {volume} {92}},\ \bibinfo
  {pages} {012323} (\bibinfo {year} {2015})}\BibitemShut {NoStop}%
\bibitem [{\citenamefont {Mezzacapo}\ \emph {et~al.}(2013)\citenamefont
  {Mezzacapo}, \citenamefont {Casanova}, \citenamefont {Lamata},\ and\
  \citenamefont {Solano}}]{Mezzacapo}%
  \BibitemOpen
  \bibfield  {author} {\bibinfo {author} {\bibfnamefont {A.}~\bibnamefont
  {Mezzacapo}}, \bibinfo {author} {\bibfnamefont {J.}~\bibnamefont {Casanova}},
  \bibinfo {author} {\bibfnamefont {L.}~\bibnamefont {Lamata}}, \ and\ \bibinfo
  {author} {\bibfnamefont {E.}~\bibnamefont {Solano}},\ }\href
  {http://stacks.iop.org/1367-2630/15/i=3/a=033005} {\bibfield  {journal}
  {\bibinfo  {journal} {New Journal of Physics}\ }\textbf {\bibinfo {volume}
  {15}},\ \bibinfo {pages} {033005} (\bibinfo {year} {2013})}\BibitemShut
  {NoStop}%
\bibitem [{\citenamefont {Mourik}\ \emph {et~al.}(2012)\citenamefont {Mourik},
  \citenamefont {Zuo}, \citenamefont {Frolov}, \citenamefont {Plissard},
  \citenamefont {Bakkers},\ and\ \citenamefont {Kouwenhoven}}]{black1}%
  \BibitemOpen
  \bibfield  {author} {\bibinfo {author} {\bibfnamefont {V.}~\bibnamefont
  {Mourik}}, \bibinfo {author} {\bibfnamefont {K.}~\bibnamefont {Zuo}},
  \bibinfo {author} {\bibfnamefont {S.~M.}\ \bibnamefont {Frolov}}, \bibinfo
  {author} {\bibfnamefont {S.~R.}\ \bibnamefont {Plissard}}, \bibinfo {author}
  {\bibfnamefont {E.~P. A.~M.}\ \bibnamefont {Bakkers}}, \ and\ \bibinfo
  {author} {\bibfnamefont {L.~P.}\ \bibnamefont {Kouwenhoven}},\ }\href
  {\doibase 10.1126/science.1222360} {\bibfield  {journal} {\bibinfo  {journal}
  {Science}\ }\textbf {\bibinfo {volume} {336}},\ \bibinfo {pages} {1003}
  (\bibinfo {year} {2012})}\BibitemShut {NoStop}%
\bibitem [{\citenamefont {Nadj-Perge}\ \emph {et~al.}(2014)\citenamefont
  {Nadj-Perge}, \citenamefont {Drozdov}, \citenamefont {Li}, \citenamefont
  {Chen}, \citenamefont {Jeon}, \citenamefont {Seo}, \citenamefont {MacDonald},
  \citenamefont {Bernevig},\ and\ \citenamefont {Yazdani}}]{black2}%
  \BibitemOpen
  \bibfield  {author} {\bibinfo {author} {\bibfnamefont {S.}~\bibnamefont
  {Nadj-Perge}}, \bibinfo {author} {\bibfnamefont {I.~K.}\ \bibnamefont
  {Drozdov}}, \bibinfo {author} {\bibfnamefont {J.}~\bibnamefont {Li}},
  \bibinfo {author} {\bibfnamefont {H.}~\bibnamefont {Chen}}, \bibinfo {author}
  {\bibfnamefont {S.}~\bibnamefont {Jeon}}, \bibinfo {author} {\bibfnamefont
  {J.}~\bibnamefont {Seo}}, \bibinfo {author} {\bibfnamefont {A.~H.}\
  \bibnamefont {MacDonald}}, \bibinfo {author} {\bibfnamefont {B.~A.}\
  \bibnamefont {Bernevig}}, \ and\ \bibinfo {author} {\bibfnamefont
  {A.}~\bibnamefont {Yazdani}},\ }\href {\doibase 10.1126/science.1259327}
  {\bibfield  {journal} {\bibinfo  {journal} {Science}\ }\textbf {\bibinfo
  {volume} {346}},\ \bibinfo {pages} {602} (\bibinfo {year}
  {2014})}\BibitemShut {NoStop}%
\bibitem [{\citenamefont {Chen}\ \emph {et~al.}(2015)\citenamefont {Chen},
  \citenamefont {Wu}, \citenamefont {Wu}, \citenamefont {Han}, \citenamefont
  {Xu}, \citenamefont {Wang}, \citenamefont {Ye}, \citenamefont {Han},
  \citenamefont {He}, \citenamefont {Cai},\ and\ \citenamefont
  {Wang}}]{wangning}%
  \BibitemOpen
  \bibfield  {author} {\bibinfo {author} {\bibfnamefont {X.}~\bibnamefont
  {Chen}}, \bibinfo {author} {\bibfnamefont {Y.}~\bibnamefont {Wu}}, \bibinfo
  {author} {\bibfnamefont {Z.}~\bibnamefont {Wu}}, \bibinfo {author}
  {\bibfnamefont {Y.}~\bibnamefont {Han}}, \bibinfo {author} {\bibfnamefont
  {S.}~\bibnamefont {Xu}}, \bibinfo {author} {\bibfnamefont {L.}~\bibnamefont
  {Wang}}, \bibinfo {author} {\bibfnamefont {W.}~\bibnamefont {Ye}}, \bibinfo
  {author} {\bibfnamefont {T.}~\bibnamefont {Han}}, \bibinfo {author}
  {\bibfnamefont {Y.}~\bibnamefont {He}}, \bibinfo {author} {\bibfnamefont
  {Y.}~\bibnamefont {Cai}}, \ and\ \bibinfo {author} {\bibfnamefont
  {N.}~\bibnamefont {Wang}},\ }\href {http://dx.doi.org/10.1038/ncomms8315}
  {\bibfield  {journal} {\bibinfo  {journal} {Nat Commun}\ }\textbf {\bibinfo
  {volume} {6}} (\bibinfo {year} {2015})}\BibitemShut {NoStop}%
\end{thebibliography}%

\end{document}